\documentclass[journal]{IEEEtran}

\usepackage{cite}
\usepackage{amsmath,amssymb,amsfonts}
\usepackage{graphicx}
\usepackage{booktabs}
\usepackage{tabularx}
\usepackage{array}
\usepackage{url}

\title{SAGEGAN: Style-Based Anomaly Detection with Gaussian Embeddings using Generative Adversarial Networks}

\author{
Thesath~Wijayasiri,
Kar~Wai~Fok,
and~Vrizlynn~L. L. Thing%
\thanks{Thesath Wijayasiri, Kar Wai Fok, and Vrizlynn L. L. Thing are with Cybersecurity Strategic Technology Center, Singapore Technologies Engineering, Singapore.}
\thanks{Corresponding author: Thesath Wijayasiri. Email: bopearachchigethesathguwantha.wijayasiri@stengg.com.}
}

\begin{document}
\maketitle

\begin{abstract}
Malware evolves faster than rule-based and signature-driven detection pipelines. This paper presents SAGEGAN, a benign-only trained malware anomaly detection framework that converts portable executable files into compact three-channel images and models benign structure through style-conditioned adversarial reconstruction. The representation combines Hilbert-mapped byte values, benign-referenced byte-transition surprise, and entropy deviation from benign software. The model encodes each image into a layer-wise style tensor aligned with a seven-stage modulated generator, rather than a single latent bottleneck. A Gaussian style prior, moment-based prior alignment, and latent consistency are used to reduce mismatch between encoded benign styles and the generator's sampled manifold. For interpretation, a deterministic encoder pathway maps each executable to a fixed style tensor, enabling repeatable layer-wise family distance, gradient sensitivity, principal component, and class-behaviour analyses. On a self-collected portable executable corpus containing malware from 214 families, the Gaussian variant achieves 89.76\% area under the receiver operating characteristic curve and 88.19\% balanced accuracy, while the genome-style variant reaches 88.03\% and 84.09\%, respectively. Without refitting model weights, benign reference statistics, or decision thresholds, the same checkpoints are evaluated on DIKE, Microsoft BIG 2015, and Lester malware subsets. The results suggest that layer-wise style modelling supports both anomaly ranking and structured post hoc analysis of how malware families depart from the benign manifold.
\end{abstract}


\section{Introduction}
Malware detection remains a moving target. Signature-based systems and handcrafted heuristics remain effective for known threats, yet they struggle against packed, obfuscated, and polymorphic binaries. Dynamic analysis can expose richer behavioural evidence, but it is computationally expensive, difficult to scale, and susceptible to sandbox evasion \cite{Souri2018Survey}. These limitations continue to motivate static learning-based detectors that can generalize beyond memorized families.

One promising line of work converts executable files into images and applies computer vision methods to the resulting representations \cite{Nataraj2011MalwareImages}. Such methods are attractive because binaries often exhibit regular structural patterns at both local and global scales. When byte streams are rendered spatially, section layout, repeated motifs, and statistical irregularities can become visually coherent and accessible to modern deep models. Recent surveys show that malware imaging has grown into an active subfield, with increasing attention to representation quality, robustness, and operational deployment concerns \cite{Bensaoud2024SurveyDL}.

Despite these advances, two persistent challenges remain. The first is the zero-day setting. In operational environments, the most important samples are often rare or previously unseen variants rather than members of well-represented classes. This weakens the assumptions behind standard supervised multi-class classification and instead motivates anomaly detection methods that model benign structure and score deviations from it \cite{Deldar2023ZeroDaySurvey}. The second challenge is interpretability. Security analysts need more than a binary output. They need models whose behaviour can be inspected, stress-tested, and related to meaningful characteristics of malicious code \cite{Saqib2024XAIMalware}.

In this work we propose style-based anomaly detection with gaussian embeddings using generative adversarial networks also known as SAGEGAN. We follow a benign-only anomaly detection formulation and make two central design choices. First, we use a compact malware image representation that preserves raw byte structure while incorporating two benign-referenced statistical cues. The aim is not to claim novelty from image construction alone, but to provide a stable and informative input space for benign manifold learning. Each binary is mapped along a Hilbert curve to preserve locality \cite{Lawder2000HilbertIndexing}. The red channel stores raw byte values, the green channel encodes byte bigram surprise under a benign reference model, and the blue channel captures deviation from the benign entropy mean.

Second, we pair this representation with a style-conditioned adversarial reconstruction framework inspired by style-based generative models \cite{Karras2020StyleGAN2}. Instead of learning only a plain auto-encoding pathway, the model uses an encoder, generator, and discriminator to learn a compact benign manifold and score departures from it through reconstruction error and discriminator response. The Gaussian variant encodes each image into a layer-wise style tensor aligned with the generator hierarchy. This tensor is regularized toward a learned Gaussian style prior through moment matching and latent consistency. A discrete genome-style variant is retained as a secondary comparison, allowing us to examine whether benign modelling is better served by an aligned continuous style space or by a discrete compositional style representation.

The central distinction of this work is that the style space is used both for anomaly scoring and for repeatable post hoc analysis. The proposed model does not treat the generator as only a reconstruction module. It also uses the generator's layer-wise modulation hierarchy as an internal coordinate system for examining how malware families differ from benign software.
The explainability provided in this work is intended to show how the model organizes and separates malware families from benign software in its learned style space, rather than to identify exact source code instructions or byte-level causes. In practical terms, it allows an analyst to inspect whether a sample appears anomalous because of broad structural differences, finer local irregularities, or family specific behaviour across the generator hierarchy. To make this analysis repeatable, we use a deterministic encoder pathway, so each executable maps to a fixed style tensor. This avoids ambiguity from random sampling during analysis and allows layer-wise distances, sensitivity measures, and family dispersion to be recomputed consistently.

We evaluate the proposed method on a self-collected PE malware corpus containing 10,820 malicious files across 214 malware families, with samples collected from the public MalwareBazaar database, together with a benign set drawn from the Lester collection \cite{MalwareBazaar,LesterDatasetKaggle}. Training is performed on benign data only. We report area under the receiver operating characteristic curve and balanced accuracy under intentionally unbalanced test mixtures, and we further test transfer to DIKE, Microsoft BIG 2015, and a disjoint Lester malware subset without refitting the learned benign statistics, model weights, or operating threshold \cite{DikeDataset,Ronen2018MicrosoftBIG,LesterDatasetKaggle}.

The main contributions of this study are as follows.

\begin{enumerate}
    \item We propose a benign-only malware anomaly detector that combines byte-aware malware imaging with a style-modulated adversarial reconstruction framework. This method is implemented through:
    \begin{itemize}
        \item a layer-wise Gaussian style inference mechanism in which the encoder maps each executable image to a seven-layer style tensor aligned with the generator hierarchy, rather than to a single latent bottleneck;
        \item moment-based prior alignment and latent consistency on generated samples, which regularize the encoded benign style distribution and reduce mismatch between the encoder's reconstruction space and the generator's sampled style manifold.
    \end{itemize}
    \item We evaluate the method on a self-collected portable executable corpus and test zero-shot transfer to DIKE, Microsoft BIG 2015, and Lester malware subsets without refitting model weights, benign reference statistics, or decision thresholds and demonstrate improvements over existing methods.
    \item We provide deterministic style-space explainability through layer-wise family distance, gradient-based layer sensitivity, principal component structure, and class-behaviour analysis, characterizing how malware families depart from the benign manifold across generator depths.
\end{enumerate}

\begin{figure}[t]
    \centering
    \includegraphics[width=\columnwidth]{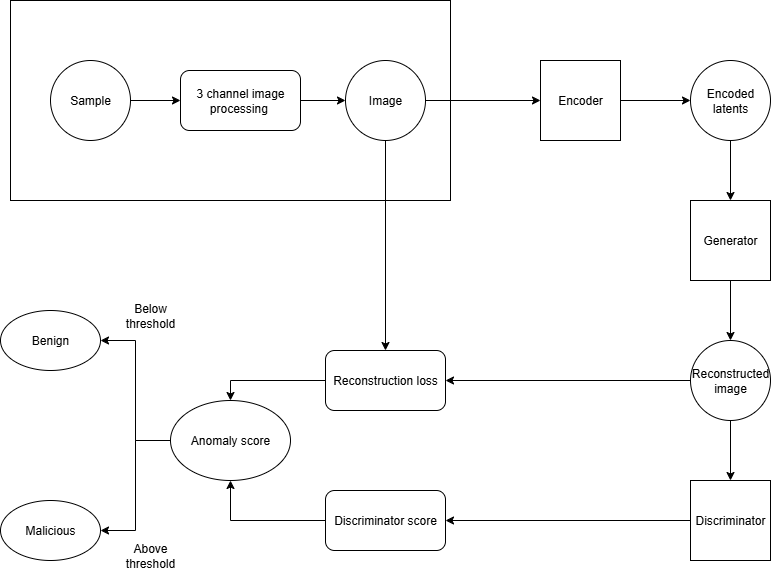}
    \caption{Overview of the SAGEGAN pipeline from binary conversion to anomaly scoring.}
    \label{fig:process}
\end{figure}

\section{Related Work}
Research on malware visualization began with byteplot-style renderings in which binaries were mapped to images and classified using texture and layout cues \cite{Nataraj2011MalwareImages}. Early work demonstrated that malware families often exhibit visually recurring structure, and later studies extended this idea through richer colour mappings and deeper discriminative backbones, particularly on Microsoft BIG and related benchmarks. Most of this literature, however, is framed as supervised family classification rather than anomaly detection.

A related strand of work focuses on improving the image representation itself. Rather than treating malware imaging as a fixed preprocessing step, these studies examine how channel design influences downstream performance. Deng et al. proposed a three-channel visualization strategy for malware classification and showed that representation choice can materially affect results \cite{Deng2023MCTVD}. Our work uses a compact three-channel representation for the same practical reason: the input should preserve meaningful structure and support stable modelling. However, the main emphasis of this paper is not the novelty of the image construction in isolation, but the way such a representation interacts with style-conditioned benign manifold learning and deterministic style-space analysis.

Anomaly detection with generative models follows a different tradition. Autoencoders reconstruct nominal data and identify anomalies through elevated residuals. Adversarial variants such as AnoGAN, EGBAD, and related methods augment reconstruction with discriminator guidance or latent inference pathways \cite{Schlegl2017AnoGAN,Zenati2018EGBAD}. These methods establish a useful recipe for image anomaly detection, but they do not directly address the statistical structure of malware bytes or the challenges of benign-only malware screening.

Recent work has begun to bridge this gap. Wijayasiri et al. introduced a consistency-constrained bidirectional adversarial framework for malware images and showed that benign-only adversarial reconstruction can produce competitive area under the curve under imbalanced conditions \cite{Wijayasiri2025CBiGAN}. Shaukat et al. used malware image features extracted from a deep backbone and trained a one-class support vector machine on benign samples, again emphasizing metrics suitable for imbalanced data \cite{Shaukat2024EAAI}. These studies are especially relevant to the present work. Compared with them, SAGEGAN differs in two ways. First, it uses a representation that combines literal byte structure with benign-referenced statistical cues. Second, it uses a style-conditioned generator and explicitly studies the deterministic style representation induced by the encoder.

Table~\ref{tab:novelty_positioning} summarizes how the proposed method is positioned relative to the closest families of malware anomaly detection approaches. The table is intended to clarify that the contribution is not simply the use of malware images or adversarial reconstruction, but the combination of benign-only screening, layer-wise style inference, and deterministic style-space analysis for family-level interpretation.

\begin{table}[t]
\centering
\caption{Positioning relative to closely related malware anomaly detection work.}
\label{tab:novelty_positioning}
\renewcommand{\arraystretch}{1.15}
\small
\begin{tabularx}{\columnwidth}{p{0.27\columnwidth}XX}
\toprule
Method family & Usual limitation & Distinction in this work \\
\midrule
Malware image classifiers & Typically supervised family classification & Benign-only anomaly screening and transfer evaluation \\
AnoGAN \cite{Schlegl2017AnoGAN} and f-AnoGAN \cite{Schlegl2019fAnoGAN} & Single latent pathway with limited malware-specific structure & Layer-wise style inference aligned to generator modulation \\
CBiGAN \cite{Carrara2021GANAutoEncoderAnomaly} and ECBiGAN \cite{Wijayasiri2025CBiGAN} & Adversarial reconstruction without style hierarchy analysis & Deterministic style-space analysis across generator depths \\
OCSVM-based malware screening & Strong anomaly framing but limited multiscale latent interpretation & Joint anomaly score and family-level style behaviour analysis \\
\bottomrule
\end{tabularx}
\end{table}

The choice of Hilbert mapping is motivated by prior work on locality-preserving space-filling curves \cite{Lawder2000HilbertIndexing}. When byte sequences are arranged through a Hilbert path rather than a naive raster scan, nearby elements in the original sequence are more likely to remain spatially adjacent in the resulting image. This is useful in executable structure, where local motifs and sectional continuity matter. We retain this principle but augment it with two benign-referenced channels. To the best of our knowledge, the combination of Hilbert-mapped bytes, benign bigram surprise, and benign entropy deviation has not been studied within a style-conditioned malware anomaly detector.

Style-based generators have transformed image synthesis by injecting latent information through modulation across multiple layers \cite{Karras2019StyleGAN,Karras2020StyleGAN2}. Their importance in the present context is not photorealistic synthesis, but the structured way in which latent information is introduced across depth. In anomaly detection, this matters because reconstruction quality alone is not sufficient; the latent space must also remain organized enough for deviations from nominal structure to be meaningfully expressed. A style-conditioned generator provides a natural mechanism for this. It imposes latent influence across multiple spatial scales while preserving a deep convolutional pathway for local detail.

This is especially attractive for malware imagery. Static binaries often contain recurring motifs at multiple spatial scales, including coarse layout regularities, medium-scale sectional patterns, and fine-grained byte-level irregularities. A style-based generator allows these scales to be conditioned through a layer-wise mechanism rather than through a single flat bottleneck. In natural image domains, style spaces are often associated with a coarse-to-fine hierarchy and with partially interpretable variation. Malware images differ substantially from natural images, so such interpretations should be treated cautiously. Even so, the architecture provides a useful operational hierarchy that can be studied for explainability.

Finally, explainability has become an increasingly important requirement in malware analysis. Recent surveys emphasize that predictive performance alone is not sufficient if model behaviour cannot be understood or audited in analyst-facing workflows \cite{Saqib2024XAIMalware,Galli2024BehavioralXAI}. This concern is particularly relevant in malware visualization, where models may latch onto superficial regularities if their representations are poorly grounded. Our work therefore treats style-space analysis as part of the method evaluation rather than as an optional add-on.

In this context, the contribution of SAGEGAN is twofold. First, it adapts style-based generative modelling to benign-only malware anomaly detection, where the objective is not family classification or image synthesis, but the modelling of benign executable structure and the detection of deviations from it. Second, it uses the resulting layer-wise style space as an explainability mechanism, allowing anomaly behaviour to be examined across generator depths through family distance, sensitivity, compactness, and principal component analyses. To the best of our knowledge, this combination of style-based adversarial anomaly detection and deterministic style-space explainability has not previously been studied for static malware screening.

\section{Methodology}
\subsection{Overview}
The proposed pipeline has two stages. First, raw binaries are converted into three-channel images that preserve local byte structure and encode benign-referenced statistical deviations. Second, these images are used to train a style-conditioned adversarial reconstruction model on benign samples only. At test time, the deterministic encoder pathway reconstructs each input and produces an anomaly score that combines reconstruction residual and discriminator response. Figure~\ref{fig:process} shows the overall pipeline.

\subsection{Binary to Image Conversion}
\begin{figure}[t]
    \centering
    \includegraphics[width=\columnwidth]{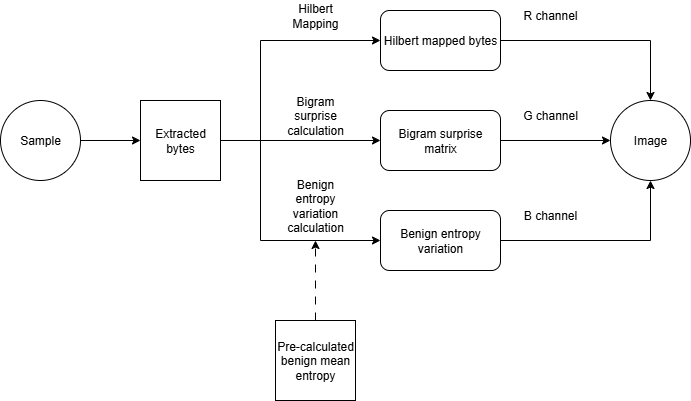}
    \caption{Binary to image conversion pipeline.}
    \label{fig:image_process}
\end{figure}

\begin{figure}[t]
    \centering
    \includegraphics[width=0.5\columnwidth]{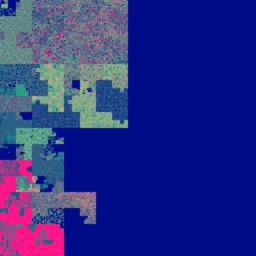}
    \caption{Example of a processed three-channel malware image.}
    \label{fig:processedimage}
\end{figure}

Each executable is parsed as a raw byte stream and mapped to a two-dimensional grid through a Hilbert space-filling curve. This yields the red channel, which preserves literal byte content while improving spatial locality relative to a simple raster layout. For files whose byte length does not match the final grid size, the byte stream is first mapped to the required Hilbert side length and then resized to $256\times256$ when necessary.

From the benign training set $\mathcal{B}$, we estimate a Laplace-smoothed byte bigram model
\[
P^{\mathcal{B}}(u \mid v)
=
\frac{C(u,v)+\alpha}{\sum_{u'} C(u',v)+256\alpha},
\]
where $C(u,v)$ counts transitions from byte $v$ to byte $u$ across benign files and $\alpha = 1$. For a file $x$ with byte sequence $b_t^{(x)}$, we define the per-position surprise score
\[
s_t^{(x)}
=
-\log_2 P^{\mathcal{B}}\bigl(b_t^{(x)} \mid b_{t-1}^{(x)}\bigr).
\]
After normalization, this forms the green channel. High values therefore correspond to local byte transitions that are unlikely under the benign reference.

For the blue channel, we compute the entropy of the file
\[
H_x
=
-\sum_{u=0}^{255} p_x(u) \log_2 p_x(u),
\qquad
p_x(u)
=
\frac{1}{N_x}\sum_{t=0}^{N_x-1} \mathbf{1}\bigl[b_t^{(x)} = u\bigr],
\]
and compare it with the benign mean entropy
\[
\bar{H}_{\mathcal{B}}
=
\frac{1}{|\mathcal{B}|}\sum_{x \in \mathcal{B}} H_x.
\]
The blue value is a scaled, centred version of $H_x - \bar{H}_{\mathcal{B}}$, so that files near the benign entropy mean map to mid-range intensity.

Together, the three channels serve a limited but important role. The red channel preserves literal structural texture, the green channel emphasizes local irregularity under a benign sequence model, and the blue channel supplies a global summary of randomness relative to benign software. The representation is intended to support stable benign manifold learning rather than to serve as the main source of contribution in isolation. Hilbert indices are cached by side length, and benign reference statistics are computed once from the benign training split and then frozen for all subsequent validation and testing.

\subsection{Model Architecture}
\subsubsection{Shared backbone}
Both model variants use the same encoder, generator, and discriminator backbone. The generator follows a style-conditioned design with seven modulated convolution blocks derived from StyleGAN2 \cite{Karras2020StyleGAN2}. Generation begins from a learned $4\times4\times512$ constant, uses nearest-neighbour upsampling, and ends with a single ToRGB head. We do not use per-layer stochastic noise injection, as preliminary experiments showed that while it improved visual sharpness, it weakened the stability of anomaly scoring. The discriminator is a strided convolutional stack trained with logistic adversarial loss and R1 regularization on real images. The encoder maps each input image to seven style vectors of dimension 512, one for each modulation stage.

This StyleGAN-inspired structure is central to the method. In a conventional encoder-decoder anomaly detector, all nominal structure is compressed through a single bottleneck and reconstructed through a decoder whose intermediate behaviour is difficult to analyse. By contrast, style-conditioned modulation distributes latent influence across multiple generator depths. This provides two advantages. First, it allows the model to reconstruct benign structure in a way that is sensitive to multiscale regularity. Second, it yields an internal hierarchy that can later be analysed for explainability. In this sense, the style architecture is not only a generator design choice but also part of the interpretive framework of the paper.

\begin{figure}[t]
    \centering
    \includegraphics[width=\columnwidth]{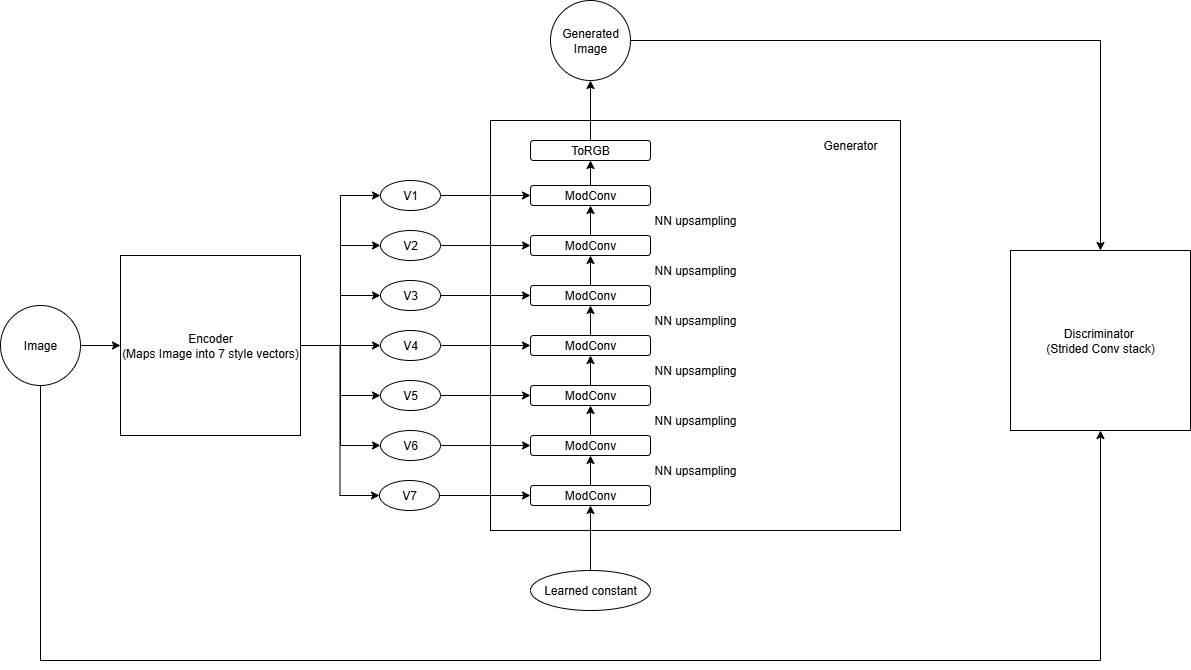}
    \caption{Encoder, generator, and discriminator architecture with style injections.}
    \label{fig:model}
\end{figure}

\subsubsection{Architecture A: layer-wise Gaussian style sampler}
The continuous variant uses a learnable Gaussian style prior with mean $\mu \in \mathbb{R}^{512}$ and diagonal standard deviation $\sigma \in \mathbb{R}^{512}$. Rather than restricting the generator to a single bottleneck vector, the model represents each image through a layer-wise style tensor
\[
W = \{w_1,\ldots,w_7\}, \qquad w_l \in \mathbb{R}^{512}.
\]
The sampler supports three style injection modes. In the broadcast mode, a single sampled style vector is shared across all generator layers. In the affine mode, a sampled style vector is transformed by layer-specific affine projections before modulation. In the independent sampling mode, each layer receives a separately sampled style vector. We achieved more consistent and stable results on the reported Gaussian model using the affine mode, which preserves a shared latent source while allowing each generator depth to receive a layer-specific style code.

The encoder maps each input image directly to a seven-layer style tensor $W_E=E(x)\in\mathbb{R}^{7\times512}$. This design differs from conventional encoder-decoder anomaly detectors, where all information is compressed through a single latent bottleneck. Here, the inferred representation is aligned with the generator's style hierarchy, making anomaly scoring and layer-wise post hoc analysis operate in the same latent space.

To reduce latent drift, the encoded benign styles are regularized toward the learned Gaussian sampler by matching their first and second moments. Let $\hat{w}$ denote the collection of encoded style vectors over a mini-batch and over all style layers. We define
\[
\mathcal{L}_{\mathrm{prior}}
=
\left\|\mu_E-\mu\right\|_2^2
+
\left\|\sigma_E^2-\sigma^2\right\|_2^2,
\]
where $\mu_E$ and $\sigma_E^2$ are the empirical mean and variance of $\hat{w}$, and $\mu$ and $\sigma^2$ are the learned sampler statistics. This term encourages benign samples encoded by $E$ to occupy the same style region used by the generator during adversarial training.

We further impose a latent consistency term on generated samples. A style tensor $W$ is sampled from the Gaussian style sampler, passed through the generator, and then re-encoded:
\[
\mathcal{L}_{\mathrm{lat}}
=
\mathbb{E}_{W}\left[\left\|E(G(W))-W\right\|_2^2\right].
\]
In implementation, the generated image is detached during this step, so the loss updates the encoder pathway and encourages it to invert the generator's style space rather than drifting to an arbitrary reconstruction-only region.

\subsubsection{Deterministic style inference for explainability}
Training uses stochastic sampling from the Gaussian style prior, but explainability is computed through the deterministic encoder pathway. For any executable image $x$, the analysis uses $W_E=E(x)$ rather than randomly sampled styles. This ensures that the same executable maps to the same style tensor every time the analysis is run, subject only to fixed model weights and deterministic preprocessing. The deterministic version is therefore central to the layer-wise distance, gradient sensitivity, principal component, and family compactness analyses reported in Section~\ref{sec:explainability}.

\subsubsection{Architecture B: genome-style latent sampler}
The second variant replaces the continuous prior with a discrete codebook. The 512-dimensional style vector is partitioned into genes, each associated with a table of learned variant embeddings. At each forward pass, one variant index is sampled per gene and the selected vectors are concatenated into a full style code, which is then supplied to the generator. In early experiments, applying the same moment alignment and latent consistency terms used in the Gaussian model degraded performance for this discrete design. We therefore use the genome-style codebook with reconstruction and adversarial training only, following the broader intuition of StyleGenes while keeping the generator and discriminator unchanged \cite{StyleGenes2022}. This makes the genome model a discrete latent counterpart within the same anomaly detection framework rather than a strictly matched optimization objective.

\subsection{Training Objectives}
Training alternates between discriminator learning, adversarial generator learning, reconstruction-based encoder-generator learning, and latent consistency learning. Let $x\sim\mathcal{B}$ denote a benign image and let $W\sim p(W)$ denote a sampled style tensor.

The discriminator is trained with binary cross-entropy on real benign images and generated images:
\[
\mathcal{L}_{D}
=
\frac{1}{2}\mathbb{E}_{x\sim\mathcal{B}}
\left[\mathrm{BCE}(D(x),1)\right]
+
\frac{1}{2}\mathbb{E}_{W\sim p(W)}
\left[\mathrm{BCE}(D(G(W)),0)\right].
\]
Lazy R1 regularization is applied to real images at a fixed interval.

The generator is trained adversarially through
\[
\mathcal{L}_{G,\mathrm{adv}}
=
\mathbb{E}_{W\sim p(W)}
\left[\mathrm{BCE}(D(G(W)),1)\right].
\]

The encoder and generator are then jointly optimized using reconstruction and prior alignment:
\[
\mathcal{L}_{E,G}
=
\lambda_{\mathrm{rec}}\left\|x-G(E(x))\right\|_1
+
\lambda_{\mathrm{prior}}\mathcal{L}_{\mathrm{prior}}.
\]
Finally, the encoder is optimized with the latent consistency loss
\[
\mathcal{L}_{E,\mathrm{lat}}
=
\lambda_{\mathrm{lat}}\mathcal{L}_{\mathrm{lat}}.
\]
This training design ensures that the generator is exposed both to sampled styles from the learned prior and to encoded benign styles from real data. The prior and latent consistency terms reduce the risk that the encoder learns a reconstruction-only latent region disconnected from the generator's adversarially trained style manifold.

We use $\lambda_{\mathrm{rec}}=10$, $\lambda_{\mathrm{prior}}=0.01$, and $\lambda_{\mathrm{lat}}=1.0$ for the Gaussian variant.

\subsection{Anomaly Scoring}
At test time, the deterministic encoder reconstructs the input image as $\hat{x}=G(E(x))$. The anomaly score combines reconstruction residual and discriminator response,
\[
s(x)
=
\alpha\,\lVert x-\hat{x}\rVert_1
-
\beta\,\ell(x),
\]
where $\ell(x)$ denotes the discriminator logit on the input. In our implementation, $\alpha=1$ and $\beta=1$, so the score is the sum of residual evidence and an inverted realism term. Higher scores indicate stronger deviation from the learned benign manifold.

The decision threshold used for balanced accuracy is selected once on a labelled validation split composed of held-out benign samples and a held-out malware validation subset that is disjoint from the final test data. The threshold is then frozen and applied unchanged to the self-collected test set and to all transfer evaluations. This threshold selection affects only the reported balanced accuracy. Area under the receiver operating characteristic curve is computed directly from continuous scores and does not depend on this operating point.

\section{Datasets and Evaluation Protocol}
\subsection{Data sources}
We study Windows portable executables converted to images using the pipeline described above. 

The study uses five PE executable datasets in total: a self-collected malware corpus, a benign executable corpus, and three public malware corpora used for external evaluation. The self-collected malware corpus contains 10,820 malicious PE files across 214 malware families, with samples collected from the public MalwareBazaar database \cite{MalwareBazaar}. The benign corpus consists of 20,000 benign executables selected from the Michael Lester collection and matched to the malware size range \cite{LesterDatasetKaggle}. The remaining malware datasets are DIKE, Microsoft Malware Classification Challenge 2015, and a disjoint Lester malware subset \cite{DikeDataset,Ronen2018MicrosoftBIG,LesterDatasetKaggle}. These datasets provide diversity across contemporary public malware samples, benchmark malware families, and independently sourced executable collections.

The main evaluation uses a self-collected malware corpus and a benign corpus drawn from a public executable collection. Transfer is then evaluated on three additional public malware datasets.

\subsubsection{Self-collected malware corpus}
We collected 10,820 malicious PE files spanning 214 malware families from MalwareBazaar \cite{MalwareBazaar}. Files were restricted to the size range of 2 MB to 50 MB to reduce trivial size effects and to keep image conversion comparable across sources.

\subsubsection{Benign corpus}
We selected 20,000 benign executables from the Michael Lester collection \cite{LesterDatasetKaggle}. The benign selection was matched to the malware size range. We also used the benign entropy statistics during preprocessing to reduce extreme covariate shift in the blue channel.

\begin{figure}[t]
    \centering
    \includegraphics[width=0.5\columnwidth]{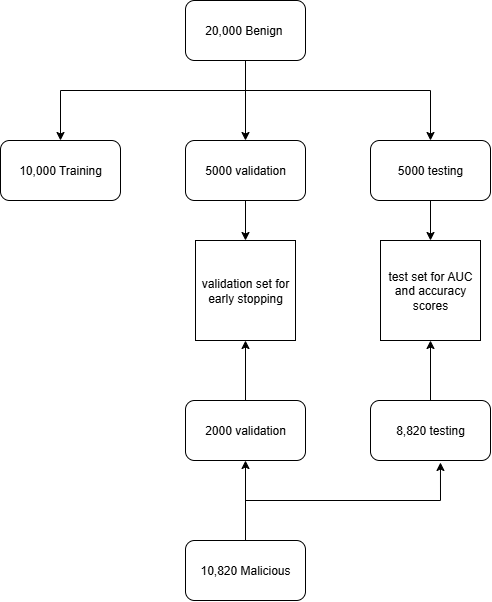}
    \caption{Primary benign data split used for training, validation, and testing.}
    \label{fig:benign_split}
\end{figure}

\subsection{Splits and leakage control}
From the benign set, 10,000 files were used for training. The remaining benign files were split evenly into validation and test partitions. Malware samples were never used to optimize the generator, encoder, discriminator, or benign reference statistics. The bigram and entropy reference values used in the green and blue channels were computed only from the benign training partition and then frozen.

For threshold selection and sanity-checked operating point analysis, we used a malware validation subset disjoint from the final malware test pool. This subset was used only to determine the single threshold applied in balanced accuracy reporting. It was not used to update model parameters or benign preprocessing statistics. Final self-collected test results use the remaining malware together with the benign test split.

\subsection{Primary and transfer test composition}
The primary self-collected test mixture evaluates the main in-domain anomaly detection setting. It combines the held out benign test set with the held out malware test set from the self-collected corpus. This experiment measures whether a model trained only on benign executables can separate unseen benign files from malware samples collected from the same overall malware source used in the main study. Because the resulting mixture is intentionally imbalanced, we report area under the receiver operating characteristic curve and balanced accuracy.

For transfer evaluation, we test whether the same learned benign manifold remains useful when the malware source changes. In each transfer experiment, the held out benign test set is kept fixed, while the malware portion is replaced with samples from one external malware dataset. We retain the same frozen benign statistics, the same trained model weights, and the same decision threshold. The transfer experiments use the following malware sources:
\begin{itemize}
    \item DIKE \cite{DikeDataset}, used to assess transfer to an external malware collection;
    \item Microsoft Malware Classification Challenge Dataset \cite{Ronen2018MicrosoftBIG}, used to assess behaviour on a widely used benchmark with established malware families;
    \item a disjoint Lester malware subset \cite{LesterDatasetKaggle}, used to assess transfer to malware samples from an independently sourced executable collection.
\end{itemize}
No model refitting, domain adaptation, preprocessing-statistic recomputation, score normalization retuning, or threshold retuning is performed on these target datasets. Thus, the transfer setting evaluates whether the anomaly score learned from benign executables remains stable across different malware sources.

\subsection{Evaluation metrics}
We report area under the receiver operating characteristic curve and balanced accuracy because the test mixtures are unbalanced and because these metrics remain meaningful under class imbalance.

Balanced accuracy is defined as
\[
\mathrm{BalAcc}
=
\frac{1}{2}(\mathrm{TPR}+\mathrm{TNR}).
\]
This metric weights the positive and negative classes equally and is therefore more informative than plain accuracy when class proportions are uneven.

The receiver operating characteristic curve plots true positive rate against false positive rate as the decision threshold varies, and the corresponding area under the curve is
\[
\mathrm{AUC}
=
\int_0^1 \mathrm{TPR}(\mathrm{FPR})\, d\,\mathrm{FPR}.
\]
AUC measures ranking quality and does not depend on the chosen class prior or the fixed threshold used for balanced accuracy.

\section{Experimental Setup}
All experiments were conducted on a single NVIDIA RTX A6000 GPU with an Intel Xeon W9 3495X CPU and 256 GB of system memory. The implementation uses PyTorch. The released training script supports longer runs, but the reported checkpoints are selected from the first 500 epochs because validation performance plateaued within this range in the observed runs. Checkpoints are selected by validation AUC, and the selected threshold is then fixed before final testing.

The generator, encoder, and discriminator were optimized with Adam using learning rate $1\times10^{-4}$ and moments $(\beta_1,\beta_2)=(0.0,0.99)$. The style sampler used a smaller learning rate of $2.5\times10^{-5}$. Inputs were resized to $256\times256$, converted to tensors, and normalized to the range implied by the image conversion pipeline.

\begin{table}[t]
\centering
\caption{Core training settings used for the reported Gaussian model.}
\label{tab:train_settings}
\renewcommand{\arraystretch}{1.15}
\small
\begin{tabularx}{\columnwidth}{lX}
\toprule
Setting & Value \\
\midrule
Epoch range used for reporting & Best checkpoint within 500 epochs \\
Image size & $256\times256$ \\
Batch size & 8 \\
Optimizer for G, E, and D & Adam, $\beta_1=0.0$, $\beta_2=0.99$ \\
Learning rate for G, E, and D & $1\times10^{-4}$ \\
Learning rate for style sampler & $2.5\times10^{-5}$ \\
Style mode & Affine per-layer styles \\
R1 regularization & $\gamma=10.0$, lazy interval 16 \\
EMA & Generator EMA, decay 0.999 \\
Mixed precision & Enabled on CUDA \\
Checkpointing & Best validation AUC \\
\bottomrule
\end{tabularx}
\end{table}

\begin{table}[t]
\centering
\caption{Implementation details for reproducibility.}
\label{tab:implementation_details}
\renewcommand{\arraystretch}{1.15}
\small
\begin{tabularx}{\columnwidth}{lX}
\toprule
Component & Setting \\
\midrule
Input & $256\times256\times3$ image \\
Style tensor & $7\times512$ \\
Generator start & Learned $4\times4\times512$ constant \\
Generator channels & 512, 512, 512, 256, 128, 64, 32 \\
Style modes implemented & Broadcast, affine, independent sample \\
Reported Gaussian mode & Affine per-layer style projection \\
Discriminator channels & 32, 64, 128, 256, 512, 512 \\
Encoder output & Seven 512-dimensional style vectors \\
Convolution kernels & $3\times3$ modulated generator convolutions; $4\times4$ strided encoder and discriminator convolutions \\
Activation & LeakyReLU with slope 0.2 \\
Anomaly score & Reconstruction residual minus discriminator logit \\
Deterministic analysis & Uses $W_E=E(x)$ with fixed model weights \\
\bottomrule
\end{tabularx}
\end{table}

Unless otherwise stated, numerical comparisons were checked across three random seeds. The Gaussian variant remained consistently ahead of the genome-style variant on the self-collected test set across these runs, although the absolute margin was modest.

\section{Results}
\subsection{Primary results on the self-collected dataset}

Table~\ref{tab:results_self} summarizes the results on the self-collected test set, which consists of 5,000 held out benign samples and 10,820 malware samples collected from the public MalwareBazaar database across 214 malware families. The Gaussian variant achieved 89.76\% AUC and 88.19\% balanced accuracy. The genome-style variant achieved 88.03\% AUC and 84.09\% balanced accuracy. The advantage of the Gaussian model is consistent with the view that explicit style prior alignment and latent consistency improve benign manifold reconstruction and anomaly separation.

\begin{table}[t]
    \centering
    \caption{Performance on the self-collected test set. Values are percentages.}
    \label{tab:results_self}
    \renewcommand{\arraystretch}{1.15}
    \small
    \begin{tabular}{lcc}
    \toprule
        Model & AUC & Balanced Accuracy \\
    \midrule
        Gaussian SAGEGAN & \textbf{89.76} & \textbf{88.19} \\
        Genome SAGEGAN & 88.03 & 84.09 \\
    \bottomrule
    \end{tabular}
\end{table}

\subsection{Comparison with additional generative baselines}
To position SAGEGAN within a broader family of benign-only generative anomaly detectors, we evaluated several additional baselines on the self-collected benchmark under the same training setting. These included Vanilla GAN, DCGAN, WGAN-GP, BigGAN, Hierarchical GAN, GANomaly, f-AnoGAN, and ECBiGAN \cite{Wijayasiri2025CBiGAN}. The results are shown in Table~\ref{tab:baseline_comparison}.

The Gaussian SAGEGAN variant achieved the highest AUC among the tested models. The genome-style variant also remained competitive and exceeded the simpler GAN-based baselines. These results suggest that adversarial generation alone is not sufficient for this task and that latent organization matters. Both SAGEGAN variants also exceeded the strongest non-SAGEGAN balanced accuracy baseline, while ECBiGAN remained the closest competing encoder-generator anomaly detector. This indicates that the gain is not merely due to adversarial reconstruction alone, but to the organization of the style space and its alignment with the encoder.

\begin{table}[t]
\centering
\caption{Comparison with generative anomaly detection baselines on the self-collected dataset. Values are percentages.}
\label{tab:baseline_comparison}
\renewcommand{\arraystretch}{1.15}
\small
\begin{tabular}{lcc}
\toprule
Model & AUC (\%) & Balanced Accuracy (\%) \\
\midrule
Gaussian SAGEGAN & \textbf{89.76} & \textbf{88.19} \\
Genome SAGEGAN & 88.03 & 84.09 \\
ECBiGAN \cite{Wijayasiri2025CBiGAN} & 86.40 & 83.10 \\
GANomaly \cite{Akcay2018GANomaly} & 84.66 & 80.88 \\
Hierarchical GAN \cite{Zhong2020HierarchicalGAN} & 80.11 & 77.71 \\
f-AnoGAN \cite{Schlegl2019fAnoGAN} & 79.41 & 76.04 \\
BigGAN \cite{Brock2019BigGAN} & 77.88 & 70.91 \\
WGAN-GP & 73.94 & 70.44 \\
DCGAN & 72.55 & 69.72 \\
Vanilla GAN & 62.49 & 63.88 \\
\bottomrule
\end{tabular}
\end{table}

\subsection{Zero-shot transfer}
Table~\ref{tab:results_transfer} reports transfer performance on DIKE, Microsoft BIG 2015, and Lester malware subsets using the same model checkpoints, frozen benign statistics, and unchanged balanced accuracy threshold.

\begin{table}[t]
    \centering
    \caption{Zero-shot transfer results. Values are percentages.}
    \label{tab:results_transfer}
    \renewcommand{\arraystretch}{1.15}
    \small
    \begin{tabular}{lccc}
    \toprule
        Model & DIKE & Microsoft & Lester \\
    \midrule
        Gaussian SAGEGAN (AUC) & \textbf{92.58} & \textbf{99.21} & \textbf{98.99} \\
        Gaussian SAGEGAN (BalAcc) & 84.52 & 96.03 & 94.24 \\
        Genome SAGEGAN (AUC) & 83.34 & 98.78 & 98.09 \\
        Genome SAGEGAN (BalAcc) & 79.80 & 95.66 & 94.01 \\
    \bottomrule
    \end{tabular}
\end{table}

The largest transfer gap appears on DIKE, where the Gaussian model exceeds the genome-style model by more than nine AUC points. On Microsoft BIG and Lester, both models perform near ceiling level, although the Gaussian variant retains a small advantage. These results suggest that the learned benign manifold retains useful anomaly ranking behaviour across the evaluated external corpora. The DIKE result is consistent with improved transfer stability from the aligned continuous latent space, although further controlled experiments are needed to establish this causally.

\subsection{Comparison with prior published methods}
Beyond the controlled same-benchmark baseline comparison above, it is also useful to position SAGEGAN relative to prior published malware anomaly detectors. Table~\ref{tab:comparison} provides contextual comparison against ECBiGAN \cite{Wijayasiri2025CBiGAN} and the OCSVM-based framework of Shaukat et al. \cite{Shaukat2024EAAI}. Because the datasets and experimental protocols are not identical across papers, these comparisons should be read as contextual rather than as strict head-to-head evaluations.

\begin{table*}[t]
\centering
\caption{Contextual comparison with prior malware anomaly detection studies. These rows do not represent fully controlled head-to-head experiments.}
\label{tab:comparison}
\renewcommand{\arraystretch}{1.2}
\small
\begin{tabularx}{\textwidth}{l l X c c}
\toprule
Method & Setting & Dataset & AUC (\%) & BA (\%) \\
\midrule
Ours & Self-collected & Self-collected PE corpus & \textbf{89.8} & \textbf{88.2} \\
CBiGAN \cite{Wijayasiri2025CBiGAN} & Self-collected & Same malware corpus with different benign setup & 86.4 & 83.1 \\
\midrule
Ours & Zero-shot transfer & Microsoft BIG 2015 without refit & \textbf{99.2} & \textbf{96.0} \\
Shaukat et al. \cite{Shaukat2024EAAI} & Mixed OCSVM & VirusShare, Malimg, and Microsoft mixture & 89.0 & 89.0 \\
\midrule
Ours & Zero-shot transfer & DIKE & \textbf{92.6} & 84.5 \\
Ours & Zero-shot transfer & Lester malware subset & \textbf{99.0} & 94.2 \\
\bottomrule
\end{tabularx}
\end{table*}

Within the controlled same-benchmark experiments, SAGEGAN compares favourably against the additional generative baselines in Table~\ref{tab:baseline_comparison}. Relative to prior published malware anomaly detectors, the transfer results appear favourable, particularly on Microsoft BIG 2015 and DIKE. However, because those literature comparisons are not protocol matched, we avoid claiming definitive superiority on that basis alone.

\subsection{Statistical validation of the anomaly score}
\begin{figure}[t]
    \centering
    \includegraphics[width=0.72\columnwidth]{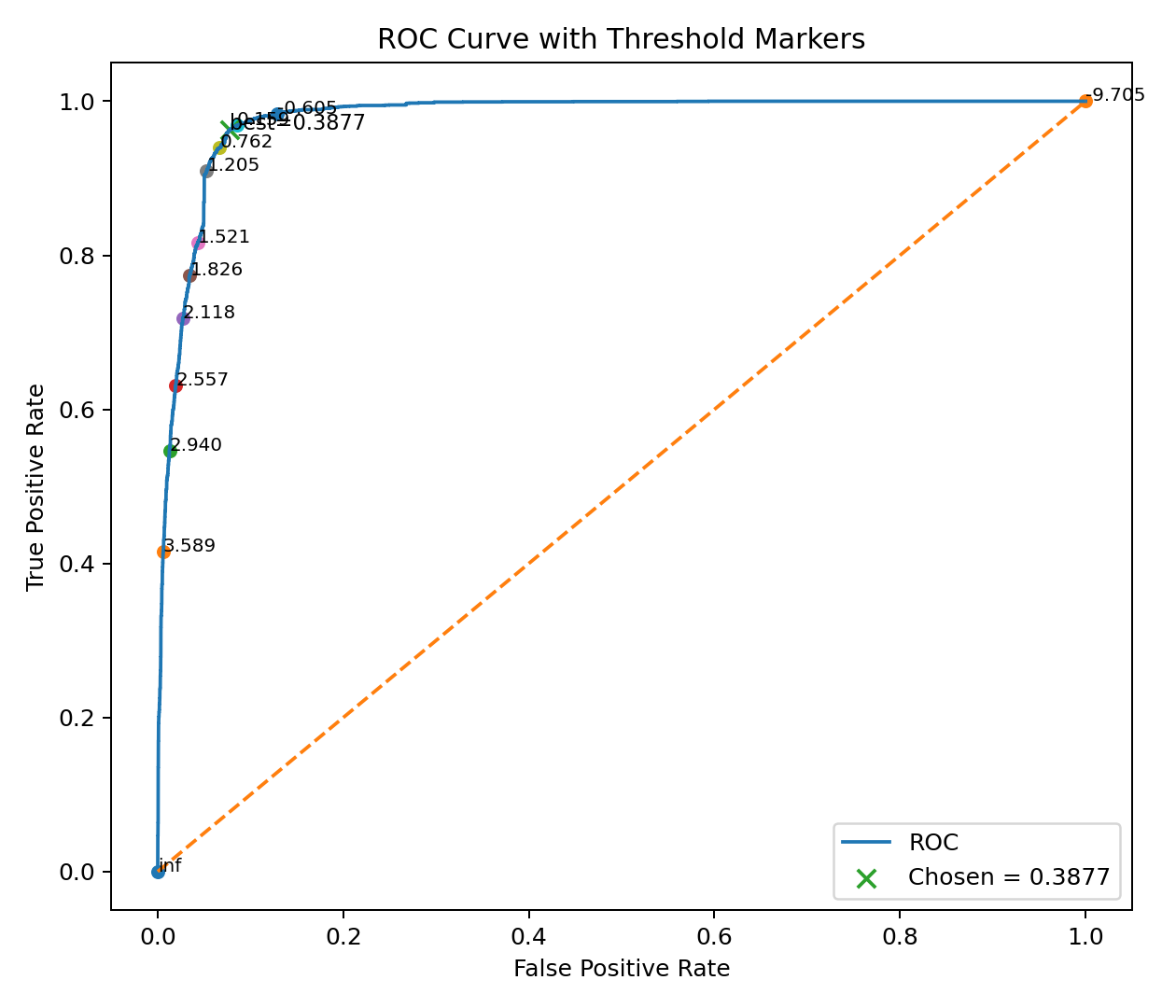}
    \caption{ROC curve with representative threshold markers on the self-collected test set.}
    \label{fig:roc_threshold_markers}
\end{figure}

\begin{figure}[t]
    \centering
    \includegraphics[width=0.72\columnwidth]{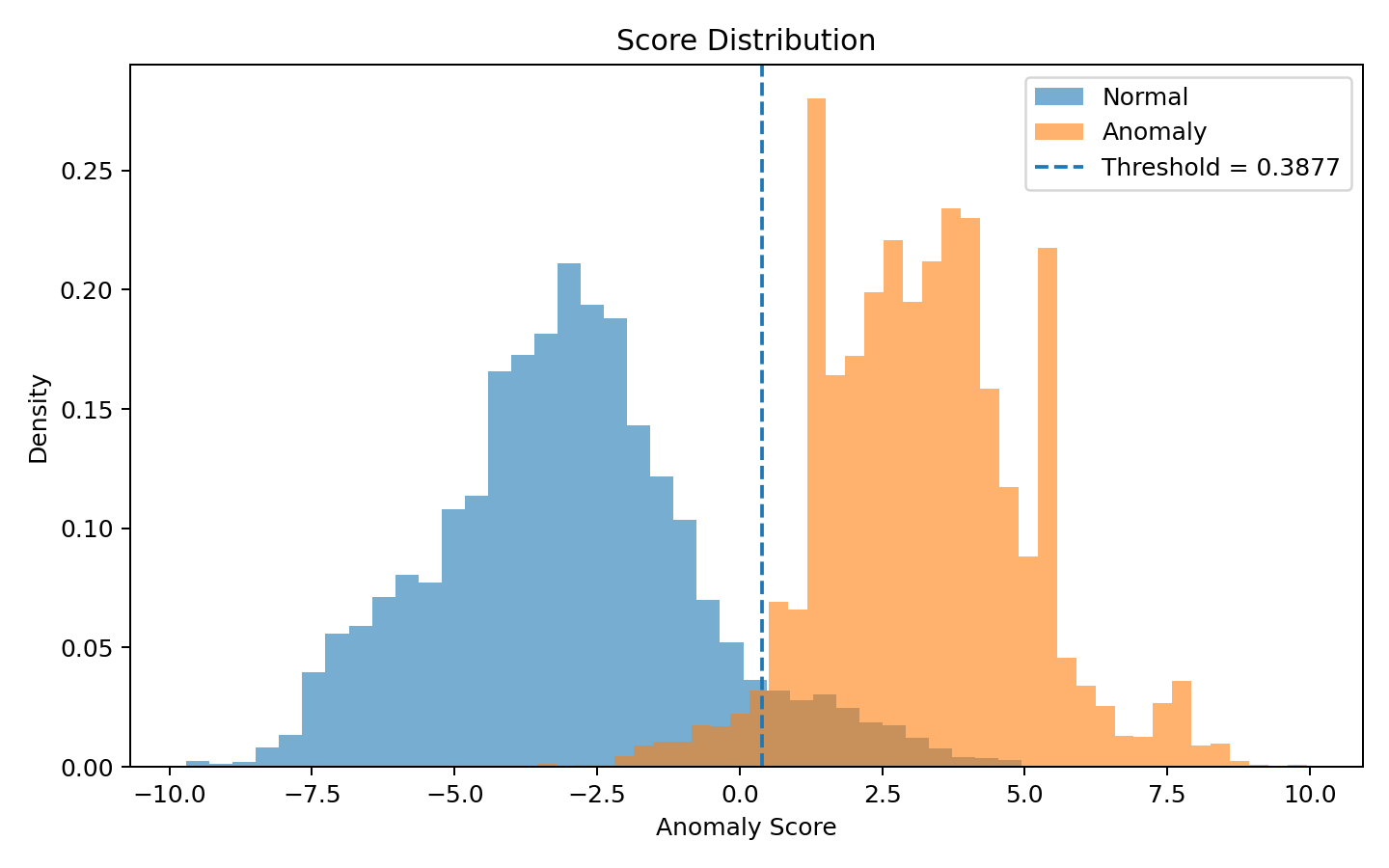}
    \caption{Anomaly score distributions for benign and malware samples on the self-collected test set.}
    \label{fig:score_histogram}
\end{figure}

\begin{figure}[t]
    \centering
    \includegraphics[width=0.74\columnwidth]{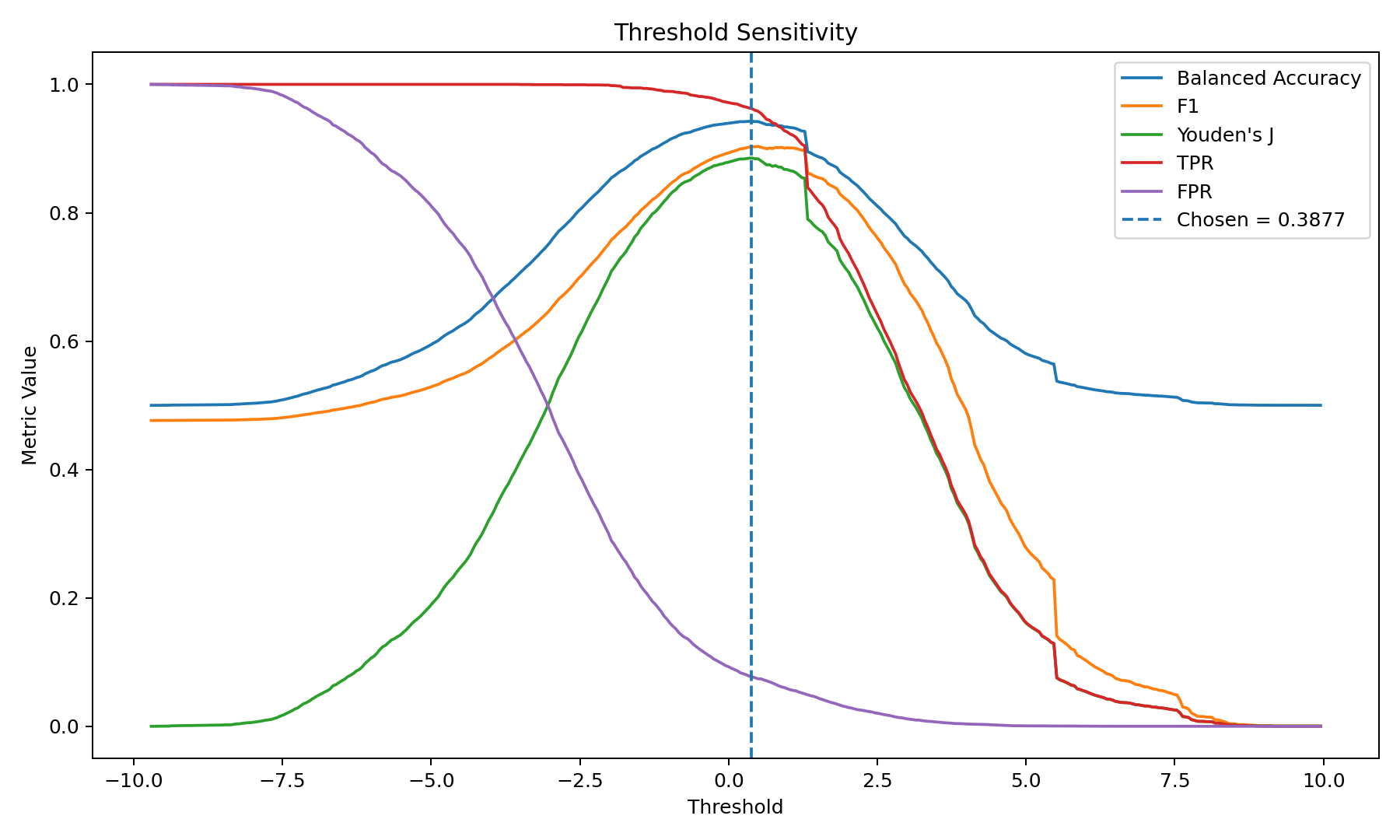}
    \caption{Threshold sensitivity of key decision metrics around the selected operating point.}
    \label{fig:threshold_sensitivity}
\end{figure}

\begin{figure}[t]
    \centering
    \includegraphics[width=0.68\columnwidth]{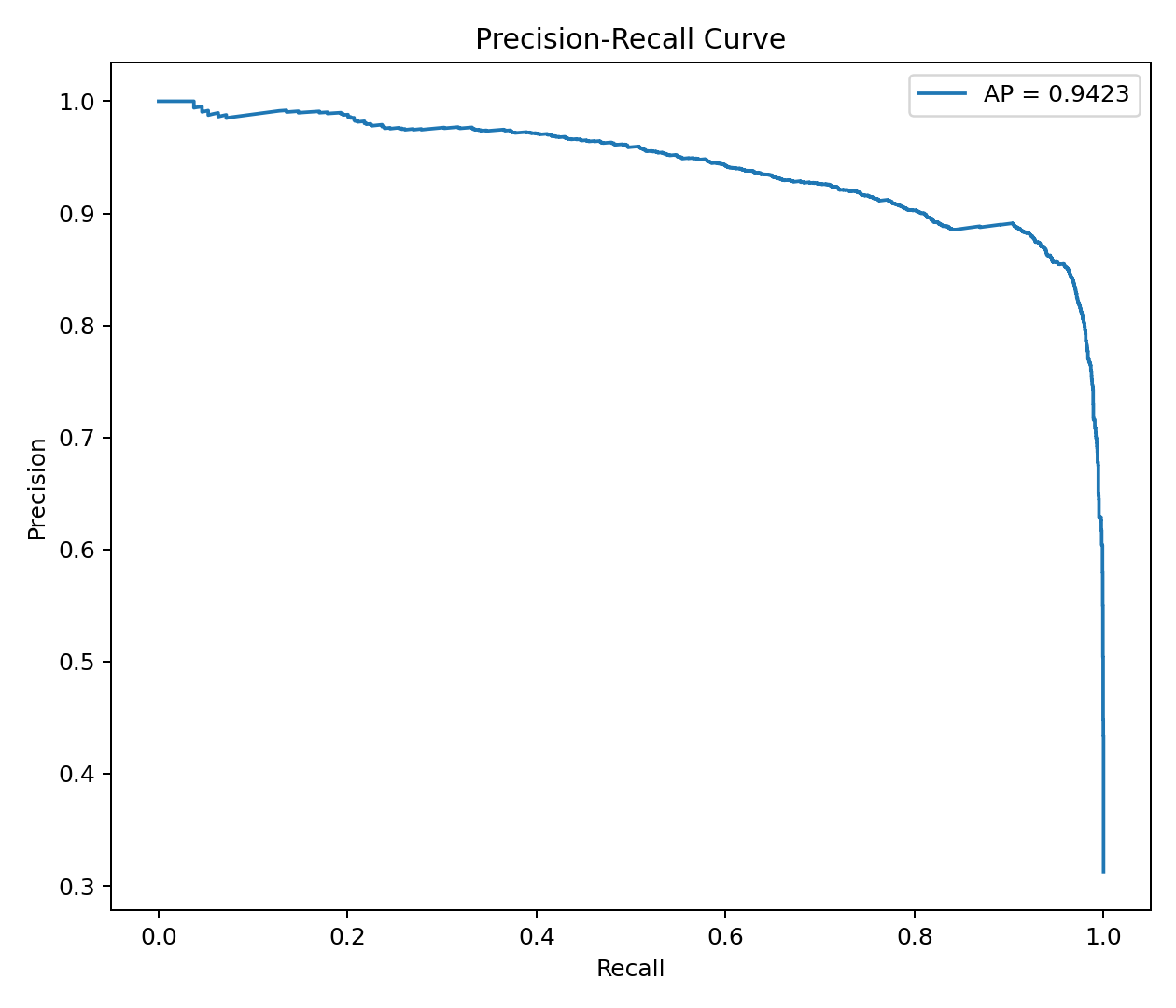}
    \caption{Precision-recall curve on the self-collected test set.}
    \label{fig:pr_curve}
\end{figure}

\begin{figure}[t]
    \centering
    \includegraphics[width=0.68\columnwidth]{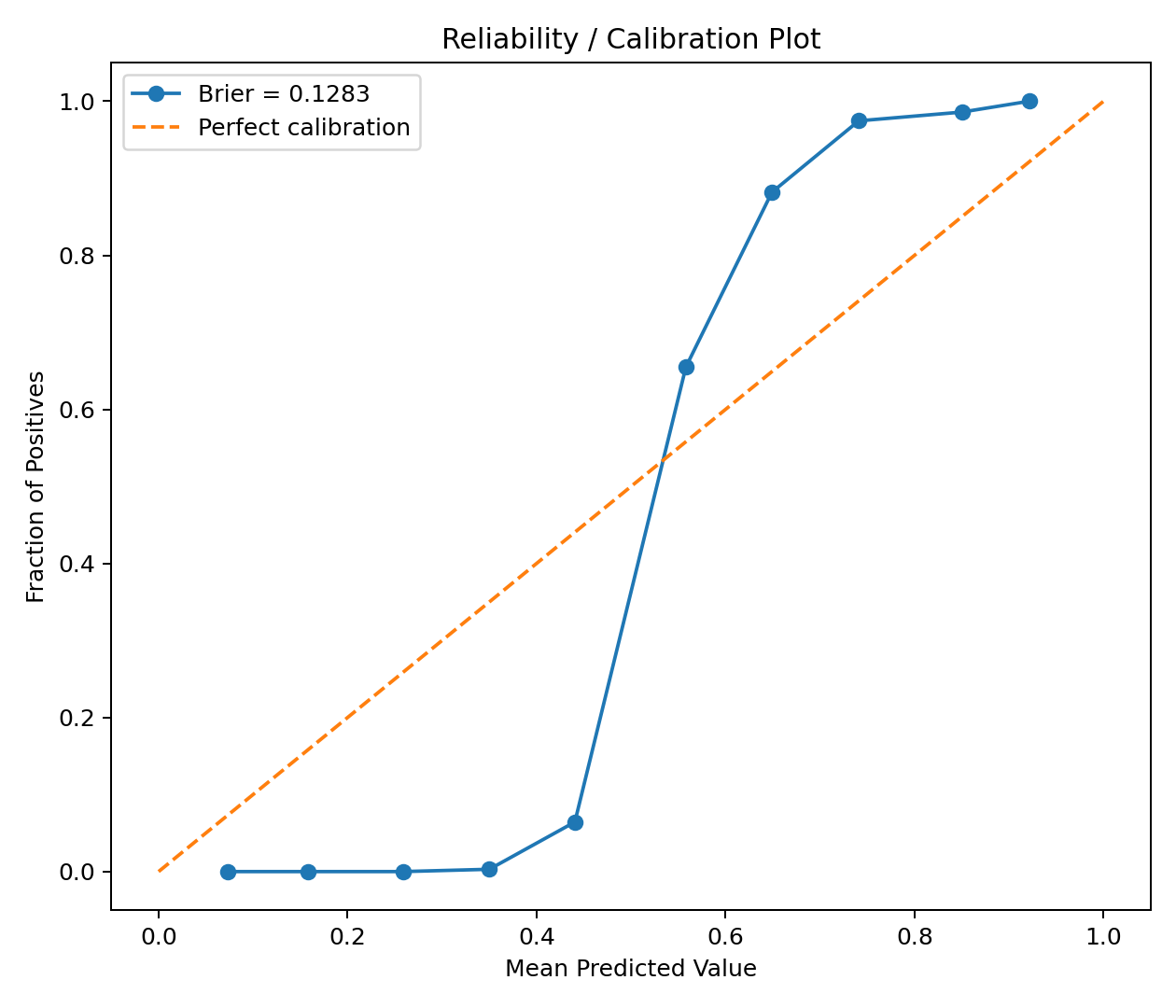}
    \caption{Reliability-style analysis of score-derived risk bins.}
    \label{fig:reliability_plot}
\end{figure}

To characterize the score beyond headline AUC and balanced accuracy, we examined the receiver operating characteristic curve, score distributions, threshold sensitivity, precision-recall behaviour, and reliability-style trends for score-derived risk bins. The receiver operating characteristic curve shows useful ranking behaviour over a broad range of operating points. The histogram of scores indicates that benign and malware samples form overlapping but meaningfully shifted distributions. Threshold sensitivity analysis shows that the selected operating point lies within a relatively stable region rather than at an isolated spike. The precision-recall curve remains informative under class imbalance, which is important for retrieval-oriented workflows. The reliability-style analysis suggests that transformed anomaly scores preserve a monotonic risk ordering, although they should not be interpreted as calibrated posterior probabilities.

\section{Discussion}
The results indicate that malware anomaly detection benefits from coupling representation design with benign manifold modelling. The proposed image pipeline preserves raw byte structure while adding two benign-referenced statistical cues. This appears to provide the generator and encoder with a target space that is both structured and informative. In particular, the red channel captures coarse layout and repeated motifs, the green channel highlights locally surprising transitions under a benign model, and the blue channel provides a coarse summary of entropy deviation.

The comparison between the Gaussian and genome-style variants suggests that latent organization matters. The Gaussian model benefits from explicit encoder alignment and latent consistency, which appear to stabilize reconstruction and improve transfer under distribution shift. The genome-style model remains competitive and offers an interesting discrete alternative, especially from an interpretability perspective, but in the present form it is best viewed as a complementary latent design rather than a stronger replacement for the continuous variant.

The emphasis on StyleGAN-type modulation is also important for interpreting the method. In this setting, style conditioning is useful not merely because it improves generation, but because it imposes a structured internal hierarchy through which benign and malicious variation can be expressed. A conventional bottlenecked decoder can reconstruct images, but it offers less direct access to how different depths contribute to family separation. Here, the deterministic style hierarchy provides a practical way to study multiscale behaviour inside the model. This is one of the main reasons the style-based design matters beyond raw benchmark performance.

The transfer results are encouraging because they were obtained without refitting model parameters, without recomputing benign statistics, and without retuning the operating threshold for each target dataset. This makes the evaluation more demanding and closer to realistic deployment conditions. At the same time, the high results on Microsoft BIG and Lester should be interpreted with caution, as these datasets may be easier than the self-collected corpus in terms of separation from contemporary benign software.

Several limitations remain. First, the study is restricted to static PE files and does not incorporate behavioural or dynamic analysis. Second, packed or heavily obfuscated binaries may alter byte-level and entropy-derived channels in ways that require separate robustness evaluation. Third, the anomaly score, while effective, is not calibrated as a true probability and should therefore be interpreted primarily as a ranking signal. Fourth, although the transfer setting is strict, the literature comparisons are not fully protocol matched. Finally, the method has not yet been evaluated under adaptive adversarial perturbations or operational constraints such as scanning latency, memory limits, and analyst triage workload.

\section{Explainability Analysis}
\label{sec:explainability}
Explainability in this work refers to post hoc analysis of the deterministic style space and its relation to malware family behaviour. It does not refer to byte-level attribution, symbolic rule extraction, or guaranteed semantic disentanglement. Instead, the goal is to understand how the model organizes benign and malicious structure internally, which parts of the style hierarchy contribute most strongly to anomaly separation, and whether malware families exhibit distinct behavioural profiles in that space.

All explainability measures are computed from the deterministic encoder output $W_E=E(x)$. This is important because stochastic style samples are useful for training the generator but unsuitable for repeatable family analysis. The deterministic pathway gives each executable a fixed representation under a fixed checkpoint, allowing layer-wise distance, gradient sensitivity, and family compactness to be recomputed without sampling ambiguity.

\subsection{Layer-wise family distance}
For each malware family $f$ and style layer $l$, we compute the distance between the family mean style vector and the benign mean style vector:
\[
d_{f,l}
=
\left\|\mu_{f,l}-\mu_{\mathcal{B},l}\right\|_2,
\]
where $\mu_{f,l}$ is the mean deterministic style vector for family $f$ at layer $l$, and $\mu_{\mathcal{B},l}$ is the corresponding benign mean. This measures where each family departs from the benign manifold across the generator hierarchy.

\subsection{Gradient-based layer sensitivity}
To estimate which style layers most influence the anomaly score, we compute a gradient-based sensitivity measure:
\[
g_{f,l}
=
\mathbb{E}_{x\in f}
\left[
\left\|\nabla_{w_l}s(x)\right\|_1
\right].
\]
This quantity measures the local sensitivity of the anomaly score to perturbations in the deterministic style vector at layer $l$. It should be interpreted as a post hoc sensitivity measure rather than as a causal explanation of the underlying binary.

\subsection{Family compactness and principal component structure}
Family compactness is measured as
\[
c_f
=
\mathbb{E}_{x\in f}
\left[
\left\|W_E(x)-\mu_f\right\|_2
\right],
\]
where $\mu_f$ is the mean flattened deterministic style tensor for family $f$. Lower values indicate that a family occupies a more compact region of style space, while higher values indicate broader within-family variation or possible subgroups. We also project flattened style tensors using principal component analysis to examine whether benign samples and malware families occupy distinguishable regions.

\subsection{Style-space analysis across malware families}
\begin{figure}[t]
    \centering
    \includegraphics[width=0.50\textwidth]{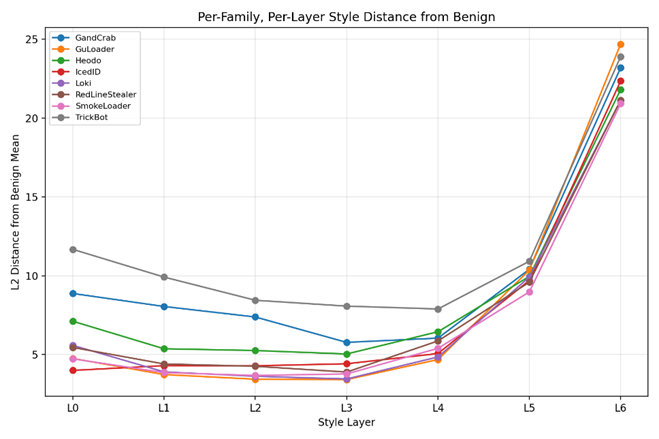}
    \caption{Per-family and per-layer style distance from the benign mean on the self-collected dataset.}
    \label{fig:self_style_distance}
\end{figure}

\begin{figure}[t]
    \centering
    \includegraphics[width=0.50\textwidth]{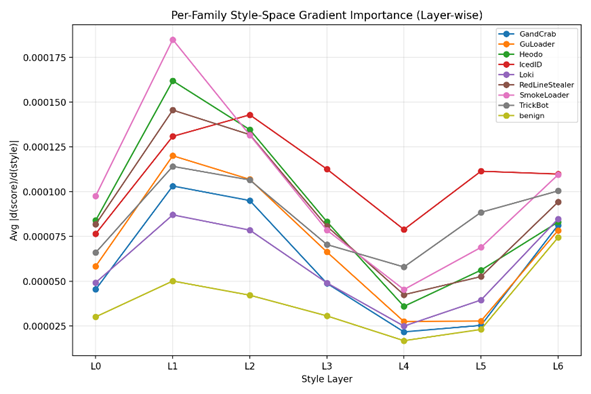}
    \caption{Per-family gradient-based layer sensitivity on the self-collected dataset.}
    \label{fig:self_style_grad}
\end{figure}

\begin{figure}[t]
    \centering
    \includegraphics[width=0.50\textwidth]{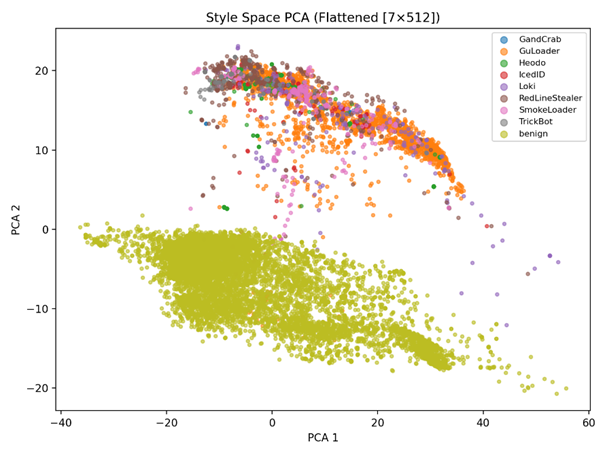}
    \caption{PCA projection of flattened deterministic style vectors on the self-collected dataset.}
    \label{fig:self_style_pca}
\end{figure}

\begin{figure}[t]
    \centering
    \includegraphics[width=0.50\textwidth]{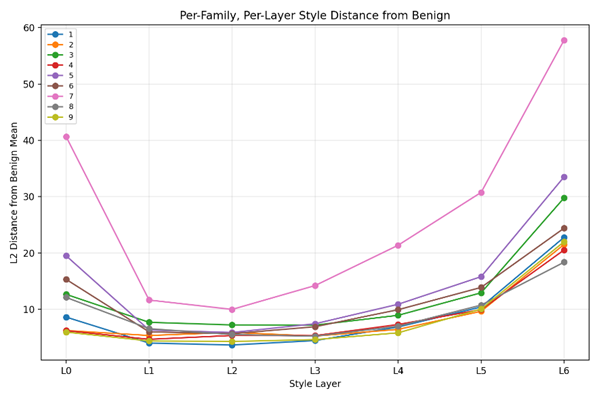}
    \caption{Per-family and per-layer style distance from the benign mean on Microsoft BIG 2015.}
    \label{fig:big_style_distance}
\end{figure}

\begin{figure}[t]
    \centering
    \includegraphics[width=0.50\textwidth]{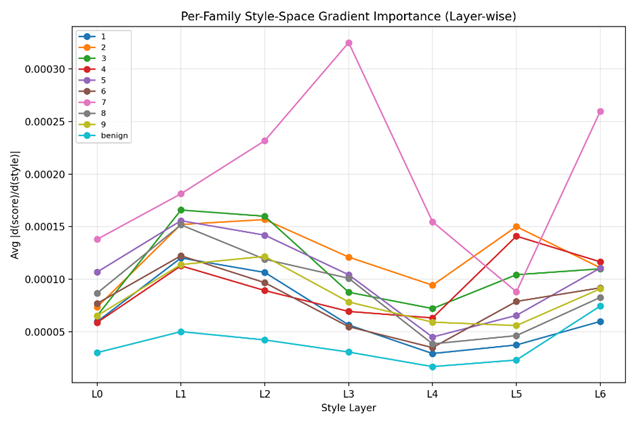}
    \caption{Per-family gradient-based layer sensitivity on Microsoft BIG 2015.}
    \label{fig:big_style_grad}
\end{figure}

\begin{figure}[t]
    \centering
    \includegraphics[width=0.50\textwidth]{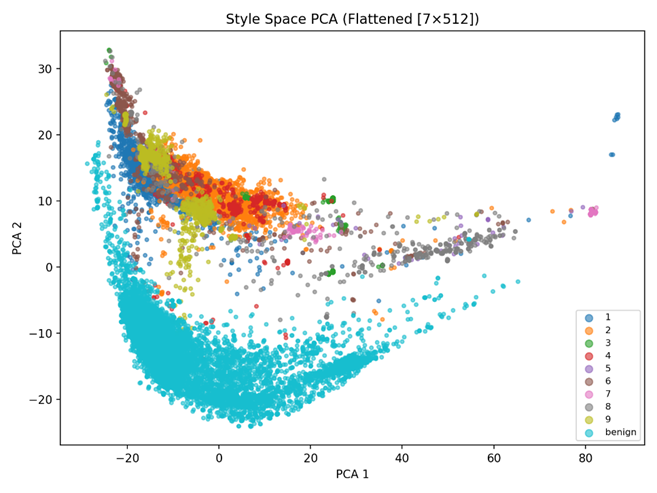}
    \caption{PCA projection of flattened deterministic style vectors on Microsoft BIG 2015.}
    \label{fig:big_style_pca}
\end{figure}

To examine how anomaly separation is distributed across the style hierarchy, we analyzed the learned deterministic style representations on both the self-collected corpus and Microsoft BIG 2015. Figures~\ref{fig:self_style_distance} to \ref{fig:big_style_pca} summarize the results.

Figure~\ref{fig:self_style_distance} shows that, on the self-collected dataset, most malware families follow a similar layer wise distance pattern: moderate separation from the benign mean in the early layers, relatively stable distances through the middle layers, and a sharp increase at the final style layer. This indicates that much of the family level separation in this dataset is expressed in the deeper generator layers, which are associated with finer structural and texture level variation. TrickBot remains more distant from benign samples across the earlier layers, suggesting a stronger coarse-to-intermediate mismatch compared with the other families.

Figure~\ref{fig:self_style_grad} shows the corresponding gradient based layer sensitivity for the self-collected dataset. The strongest sensitivity is generally observed around the early and middle style layers, particularly around L1 and L2, while sensitivity decreases around L4 before rising again for several families at L5 and L6. This suggests that the anomaly score is not controlled only by the layer with the largest distance from benign. Instead, different layers contribute differently: some layers encode large deviations, while others have greater local influence on the final anomaly score.

Figure~\ref{fig:self_style_pca} shows that benign samples occupy a broad but clearly separated region of the flattened deterministic style space. The malware families form a distinct upper cluster with partial family level organization, rather than collapsing into a single undifferentiated anomalous group. This supports the view that the deterministic encoder preserves meaningful family structure while still separating malware from benign software.

Figure~\ref{fig:big_style_distance} presents the same layer-wise distance analysis for Microsoft BIG 2015. The separation pattern is more heterogeneous than in the self-collected dataset. Several families show increasing distance toward deeper layers, but the 7th family, Kelihos ver1, exhibits substantially larger distances across the hierarchy, especially at L0 and L6. This indicates that the Microsoft BIG families do not depart from the benign manifold in a uniform way; some are separated mainly through late layer differences, while others show strong multiscale deviation.

Figure~\ref{fig:big_style_grad} further supports this observation. The gradient sensitivity profiles differ noticeably across Microsoft BIG families, with some families showing pronounced peaks in the middle layers and others showing stronger sensitivity in the later layers. The benign curve remains comparatively lower, indicating that the anomaly score is more sensitive to style perturbations for malware samples than for benign samples. This provides additional evidence that the style hierarchy captures family dependent anomaly behaviour rather than a single shared malware pattern.

Finally, Figure~\ref{fig:big_style_pca} shows that the Microsoft BIG malware samples occupy structured regions that are separated from the benign cluster, while also showing greater overlap and spread among malware families than in the self-collected PCA projection. This suggests that the learned style space preserves both benign-malware separation and within-malware variation. Taken together, Figures~\ref{fig:self_style_distance}--\ref{fig:big_style_pca} show that malware families do not separate from benign uniformly across the style stack. Instead, the model captures different class-behaviour profiles, including late-layer dominant separation, multiscale deviation, compact family clusters, and more diffuse family structure.

The PCA projections further indicate that benign samples occupy a distinct region of style space, while malware samples lie outside that region yet retain family-level organization. In other words, the encoder does not simply collapse all malware into a single off-manifold cluster. Instead, the learned deterministic style representation preserves meaningful internal structure while separating benign from malicious samples.

Table~\ref{tab:class_behaviour} summarizes these observed class behaviour patterns in the deterministic style space. The table links the empirical evidence from the layer-wise distance, gradient sensitivity, and PCA analyses to the corresponding interpretation of how different malware families depart from the benign manifold.

\begin{table}[t]
\centering
\caption{Summary of class-behaviour patterns in deterministic style space.}
\label{tab:class_behaviour}
\renewcommand{\arraystretch}{1.15}
\small
\begin{tabularx}{\columnwidth}{XXX}
\toprule
Pattern & Style-space evidence & Interpretation \\
\midrule
Late-layer dominant & Low to moderate $d_{f,l}$ at early layers and larger $d_{f,l}$ at deeper layers & Fine-grained byte or local texture irregularity dominates \\
Multi-scale deviation & Elevated $d_{f,l}$ across several layers & Coarse layout and fine structure differ from benign software \\
Compact family & Low $c_f$ and compact PCA region & Internally consistent family behaviour \\
Diffuse family & High $c_f$ or dispersed PCA region & Within-family variation or possible subgroups \\
\bottomrule
\end{tabularx}
\end{table}

\subsection{Class behavioural patterns}
A more precise form of explainability comes from examining class behaviour rather than only aggregate family separation. Here, class behaviour refers to the way a malware family expresses itself across the style hierarchy through its distance profile, sensitivity profile, and dispersion in the projected style space. This allows us to ask not merely whether a family is separable from benign software, but how that separation is distributed and how consistent the family remains internally.

Some families exhibit a predominantly late-layer pattern, with modest deviation in shallow layers and substantially stronger deviation in deeper layers. This suggests that their anomaly signal is driven more by fine-grained statistical irregularity than by broad layout mismatch. Other families remain elevated across a larger portion of the hierarchy, indicating a multiscale deviation profile in which both coarse and fine structure differ from benign software. Families that appear compact in PCA space suggest stronger internal regularity, whereas more diffuse families indicate broader within-family variation or the presence of structurally distinct subgroups.

This class-behaviour view is useful because it shifts explainability away from generic statements about the model and toward family-specific interpretation. A detector may achieve high AUC while relying on different internal evidence for different classes. By examining layer-wise behaviour and family compactness, we gain a clearer picture of whether a class is separated by broad structural mismatch, local statistical texture, or a combination of the two. This does not constitute causal explanation in the strict sense, but it does provide a structured account of how family-level behaviour is represented by the model.

\subsection{Scope of the explainability claims}
The explainability results support three concrete claims. First, the learned deterministic style space preserves family-level organization rather than collapsing all malware into a single anomalous mass. Second, anomaly separation is distributed differently across the style hierarchy for different families, indicating that malware does not violate the benign manifold in one universal way. Third, class-behaviour patterns in style space provide a practical basis for post hoc interpretation and error analysis.

At the same time, the analysis has clear boundaries. We do not claim that individual style layers have fixed semantic labels, nor do we claim that the observed behavioural profiles prove causal mechanisms in the underlying binaries. The contribution is more specific: the style-conditioned architecture yields an internal representation whose family-level structure can be analysed in a disciplined and repeatable way, providing a useful form of post hoc explainability for malware anomaly detection.

\section{Conclusion}
We presented SAGEGAN, a benign-only malware anomaly detection framework that combines a compact byte-aware image representation with style-conditioned adversarial reconstruction. The proposed representation uses Hilbert-mapped raw bytes together with benign-referenced bigram surprise and entropy deviation, while the main modelling emphasis lies in the StyleGAN-inspired generator and the structured latent hierarchy it provides. Within this framework, a Gaussian layer-wise style latent with moment-based encoder alignment achieves the strongest overall performance on the self-collected benchmark and transfers effectively to DIKE, Microsoft BIG 2015, and Lester malware subsets without refitting.

A second contribution is the deterministic explainability analysis of the learned style space. Rather than treating explainability as generic attribution, we focus on family-level structure, layer-wise deviation from benign software, gradient-based layer sensitivity, and class-behaviour patterns across the style hierarchy. The results suggest that malware families differ not only in whether they are anomalous, but also in how that anomaly is distributed across spatial scales and how coherently each family is organized in the latent space.

Future work will extend this framework to additional executable formats, incorporate dynamic traces, examine more explicit gene-aware scoring for the discrete model, and study robustness under packing, obfuscation, and adversarial perturbation.

\bibliographystyle{IEEEtran}
\bibliography{references}

\end{document}